\documentclass[prx,aps,letterpaper,twocolumn,accepted=2018-16-04]{quantumarticle}

\usepackage{bbm}
\usepackage{overpic}
\usepackage{amsmath}
\usepackage{epsfig,psfrag}
\usepackage{pstricks}
\usepackage{dcolumn}
\usepackage{bm}
\usepackage{graphicx}
\usepackage{color}
\usepackage{version}
\usepackage{hyperref}
\usepackage{natbib}
\usepackage{multicol}

\usepackage{float}

\usepackage{xcolor}
\usepackage{braket}

\begin{document}

\date{\vspace{-5ex}}

\title{Lattice Surgery with a Twist: \\ Simplifying Clifford Gates of Surface Codes}
\author{Daniel Litinski and Felix von Oppen}
\affiliation{Dahlem Center for Complex Quantum Systems and Fachbereich Physik, Freie Universit\"at Berlin, Arnimallee 14, 14195 Berlin, Germany}

\begin{abstract}

We present a planar surface-code-based scheme for fault-tolerant quantum computation 
which eliminates the time overhead of single-qubit Clifford gates, and implements long-range multi-target CNOT gates with a time overhead that scales only logarithmically with the control-target separation. This is done by replacing hardware operations for single-qubit Clifford gates with a classical tracking protocol. Inter-qubit communication is added via a modified lattice surgery protocol that employs twist defects of the surface code. The long-range multi-target CNOT gates facilitate magic state distillation, which renders our scheme fault-tolerant and universal.

\end{abstract}

\maketitle

\section{Introduction}\label{sec:intro}

The performance of quantum computers is limited by the coherence times of the underlying physical qubits. Quantum error correction~\cite{Preskill1998} offers the possibility to enhance the qubits' survival times by encoding quantum information using logical qubits consisting of many physical qubits. Topological quantum error-correcting codes~\cite{Kitaev2003,TerhalRMP} are of particular interest, as they only require the measurement of spatially local operators~--~a feature that is compatible with the local operations accessible in two-dimensional solid-state qubit architectures, such as superconducting qubits~\cite{Devoret2013}, spin qubits~\cite{Loss1998}, or Majorana-based qubits~\cite{Lutchyn2017}.

Quantum error-correcting codes typically operate in cycles. In each code cycle, mutually commuting operators called stabilizers~\cite{Gottesman1997} are measured to reveal the error syndrome, which is used to determine and correct errors. Surface codes~\cite{Bravyi1998,Campbell2016} are topological codes that feature a high error threshold~\cite{Wang2010,Andrist2016}, and only require the measurement of four-qubit stabilizer operators for the readout of the error syndrome. The low-weight stabilizers are an advantage over other codes such as color codes~\cite{Bombin2006,Landahl2011}, which require the measurement of six-qubit operators. This facilitates  syndrome readout in many physical architectures such as superconducting qubits, where the measurement of higher-weight stabilizers requires more potentially faulty controlled-not (CNOT) gates.

\begin{figure}[b!]
\def\svgwidth{\linewidth}
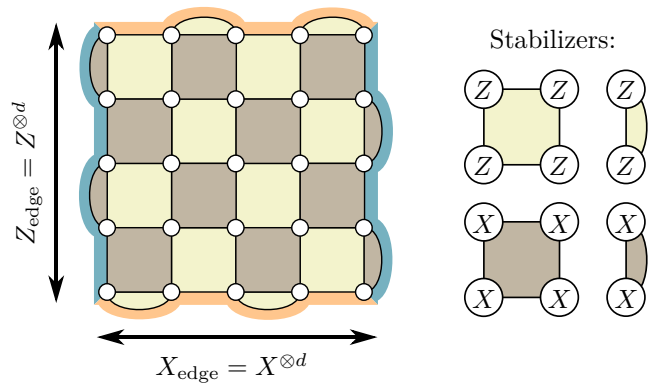
\caption{An example of a surface code qubit with code distance $d=5$. Physical qubits are located on the vertices, and the faces define the two- and four-qubit $Z$ type (bright) and $X$ type (dark) stabilizer operators. $X$ strings along the $X$ edge (orange) are logical $X_L$ operators, whereas $Z$ strings along $Z$ edges (blue) are $Z_L$ operators.}
\label{fig:qubit}
\end{figure}

\begin{figure*}[t!]
\centering
\def\svgwidth{\linewidth}
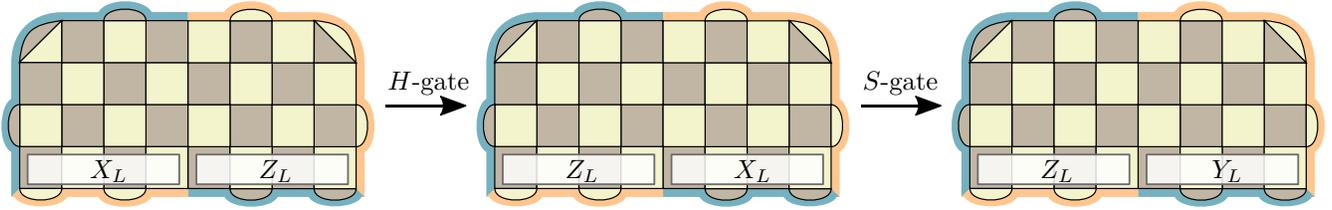
\caption{An example of edge tracking with a wide surface code qubit. Starting from the default encoding $X_{\rm edge} = X_L$ and $Z_{\rm edge} = Z_L$, an $H$ gate changes it to $X_{\rm edge} = Z_L$ and $Z_{\rm edge} = X_L$, and a subsequent $S$ gate modifies it to $X_{\rm edge} = Z_L$ and $Z_{\rm edge} = Y_L$.}
\label{fig:edgetracking}
\end{figure*}

The main drawback of surface codes in comparison to color codes is the absence of transversal single-qubit Clifford gates, i.e., the gates that are products of the Hadamard gate $H$ and the phase gate $S$. While the transversal Clifford gates of color codes provide them with fast logical $H$ and $S$ gates, defect-based proposals for surface codes~\cite{Fowler2012} implement the $H$ gate via a multi-step measurement protocol, and the $S$ gate via a distilled ancilla qubit.
In order to lower the overhead of single-qubit Clifford gates, surface code qubits can be encoded using twist defects~\cite{Bombin2010}, which are essentially Majoranas that can be braided via code deformation~\cite{Brown2017}. It was pointed out that braiding of twists can also be implemented via a classical tracking protocol~\cite{Hastings2015}, in accordance with the Gottesman-Knill theorem~\cite{Gottesman1999}.

In this work, we present a scheme that implements this tracking protocol for planar surface codes, as opposed to twist-based encodings. We refer to this protocol as \textit{edge tracking}. 
In our scheme, Clifford completeness is achieved via a modified lattice surgery~\cite{Horsman2012} protocol. Twist defects are no longer used to encode quantum information, but reappear in lattice-surgery protocols involving the logical $Y_L$ operator, so that we refer to the protocol as \textit{twist-based lattice surgery}. Our scheme provides long-range multi-target CNOT gates~--~i.e., CNOTs with one control and arbitrarily many targets~--~between any set of edge-tracked surface code qubits. These gates are particularly useful for magic state distillation~\cite{Bravyi2005}, which completes the universal gate set by fault-tolerantly implementing the $T$ gate (or $\pi/8$ gate). Our scheme not only eliminates the need for hardware operations for single-qubit Clifford gates, but also conceptually simplifies the twist-defect-based approach to surface-code quantum computing. Even though our scheme features twist defects and dislocation lines, the only concepts necessary to understand our scheme are the encoding of logical qubits and the measurement of two-qubit parity operators.  We discuss the implementation of the single-qubit Clifford gates, CNOT gates, and $T$ gates in Secs.~\ref{sec:edgetracking}, \ref{sec:surgery} and \ref{sec:2d}, respectively. In a concluding section, we discuss our scheme in the context of possible hardware implementations and in comparison to alternative topological codes.

\section{Edge Tracking}
\label{sec:edgetracking}

The basic framework of our scheme are physical qubits arranged on a 2D square lattice which allow for the measurement of local stabilizer operators. Examples of possible physical realizations include superconducting qubits emulating stabilizer measurements using ancilla qubits and CNOT gates~\cite{Fowler2012}, or Majorana-based qubits using direct measurements of the stabilizers via Majorana fermion parity measurements~\cite{Karzig2016}. A single surface code qubit can be defined using the checkered square shown in Fig.~\ref{fig:qubit}, where physical qubits are located at the vertices. We refer to the Pauli operators of the physical qubits as $X$, $Y$, and $Z$. The faces define the $X^{\otimes n}$- and $Z^{\otimes n}$-stabilizers of the code, where $n$ is the number of qubits that are part of the face. The figure shows an example of a code with code distance $d=5$, but this construction can be generalized to arbitrary code distances.

Surface code qubits have two distinct types of boundaries, usually referred to as rough and smooth edges. Here, we call them $X$ and $Z$ edges in analogy to the logical Pauli operators $X_L$ and $Z_L$ that they encode. Surface code qubits can be easily initialized in the logical $+1$-eigenstates $\ket{0_L}$ and $\ket{+_L}$ of $Z_L$ and $X_L$ by initializing all physical qubits in the corresponding physical states $\ket{0}$ and $\ket{+}$, measuring all stabilizers, and correcting the errors. Conversely, they can be read out in the $X_L$ and $Z_L$ basis by measuring all physical qubits in the $X$ or $Z$ basis, and performing classical error correction.

We define the operator $X_{\rm edge}$  ($Z_{\rm edge}$) as the string of $X$ operators ($Z$ operators) on all physical qubits along an $X$ edge ($Z$ edge). In the default encoding, $X_{\rm edge} = X_L$ and $Z_{\rm edge} = Z_L$.
The edge tracking procedure that we now introduce essentially modifies which logical operators are encoded by $X_{\rm edge}$ and $Z_{\rm edge}$.
Logical single-qubit Clifford gates map the logical Pauli operators $X_L$, $Y_L$, and $Z_L$ onto other Pauli operators. In particular, an $H$ gate maps $X_L \rightarrow Z_L$, $Y_L \rightarrow -Y_L$, and $Z_L \rightarrow X_L$. An $S$ gate maps $X_L \rightarrow Y_L$, $Y_L \rightarrow -X_L$, and $Z_L \rightarrow Z_L$. Thus, we can replace single-qubit Clifford gates by a classical tracking procedure. This is essentially the content of the Gottesman-Knill theorem~\cite{Gottesman1999}, which states that Clifford gates can be simulated efficiently on a classical computer. For now, we only consider tracking of single-qubit Clifford gates $H$ and $S$, whereas CNOT gates are performed explicitly.

In order to combine this tracking scheme with lattice surgery, it will be convenient to use the wide qubits shown in Fig.~\ref{fig:edgetracking} instead of the square qubits that were previously introduced. These qubits have an $X$ and $Z$ edge on the same side, such that the logical operators $X_L$, $Y_L$ and $Z_L$ can all be accessed by lattice surgery from the same side of the qubit. Compared to square qubits with the same code distance, this comes at the price of a larger number of physical qubits for each logical qubit. The figure also shows an example of edge tracking. The default encoding is $X_{\rm edge} = X_L$ and $Z_{\rm edge} = Z_L$. An $H$ gate changes the encoding to $X_{\rm edge} = Z_L$ and \linebreak $Z_{\rm edge} = X_L$. A subsequent $S$ gate modifies it to $X_{\rm edge} = Z_L$ and $Z_{\rm edge} = Y_L$.

\section{Lattice surgery with a twist}
\label{sec:surgery}

\begin{figure}[b!]
\centering
\def\svgwidth{\linewidth}
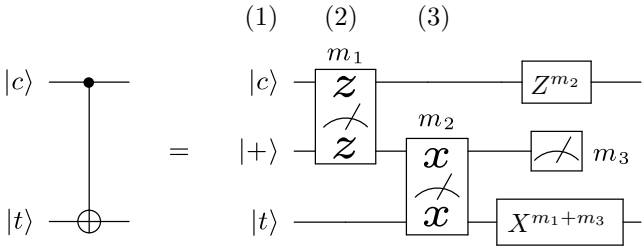
\caption{A CNOT between a control $\ket{c}$ and a target $\ket{t}$ is equivalent to a $Z\otimes Z$ parity measurement between $\ket{c}$ and an ancilla in the $\ket{+}$ state, followed by an $X\otimes X$ parity measurement between ancilla and $\ket{t}$, and finally a $Z$ basis measurement of the ancilla. The measurement outcomes determine a Pauli correction.}
\label{fig:cnotcircuit}
\end{figure}

Edge tracking requires a suitable CNOT gate protocol in order to be useful for universal quantum computation. This is provided by twist-based lattice surgery. It essentially implements the circuit identity shown in Fig.~\ref{fig:cnotcircuit} for edge-tracked qubits. Here, a CNOT between a control and target qubit corresponds to three measurements: a $Z \otimes Z$ parity measurement between the control and an ancilla initialized in the $X$ eigenstate $\ket{+}$, a subsequent $X \otimes X$ parity measurement between ancilla and target, and a final $Z$ basis readout of the ancilla qubit. In order to use this protocol for \textit{logical} CNOTs, measurements of logical two-qubit parity operators are required, e.g., operators such as $Z_L \otimes Z_L$, which are nonlocal operators involving $2d$ physical qubits.

\begin{figure}[b!]
\centering
\def\svgwidth{\linewidth}
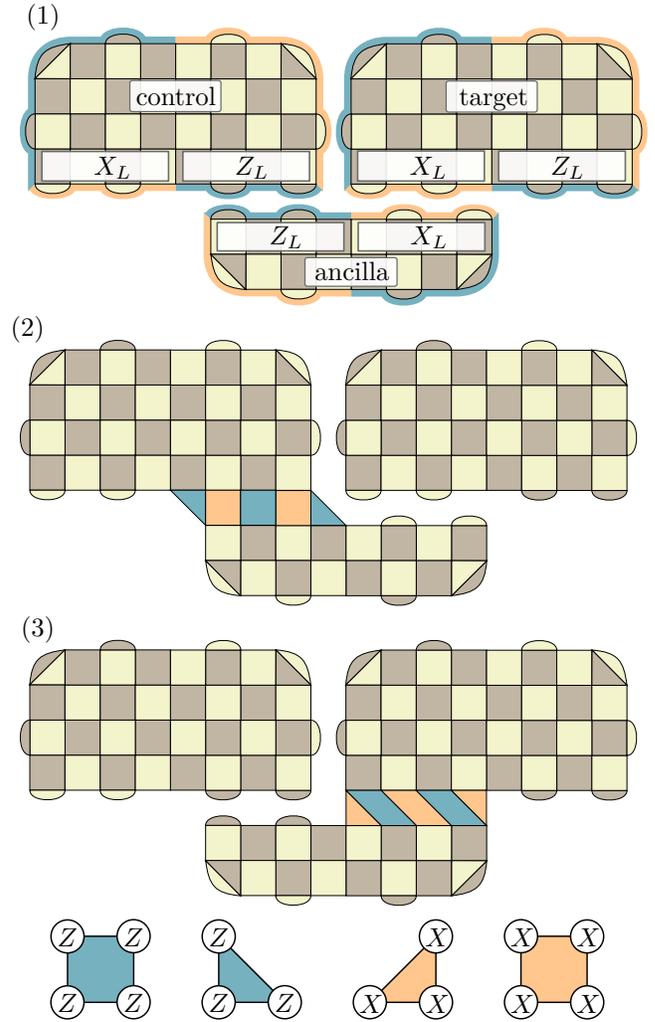
\caption{CNOT by lattice surgery corresponding to the gate circuit in Fig.~\ref{fig:cnotcircuit}. $(1)$ All qubits are in the default encoding $X_{\rm edge} = X_L$ and $Z_{\rm edge} = Z_L$, and the ancilla is initialized in the $\ket{+}$ state. $(2)$ To measure the $Z_L \otimes Z_L$ parity between control and target, the two-qubit boundary stabilizers are merged (orange), and new $Z$ type stabilizers (blue) are introduced, whose product is precisely the parity. $(3)$ Similarly, the $X_L \otimes X_L$ parity between ancilla and target is measured by the product of new $X$ type stabilizers (orange). }
\label{fig:neighborcnot}
\end{figure}

\subsection{Nearest-neighbor CNOT}

\begin{figure*}[t!]
\centering
\def\svgwidth{\linewidth}
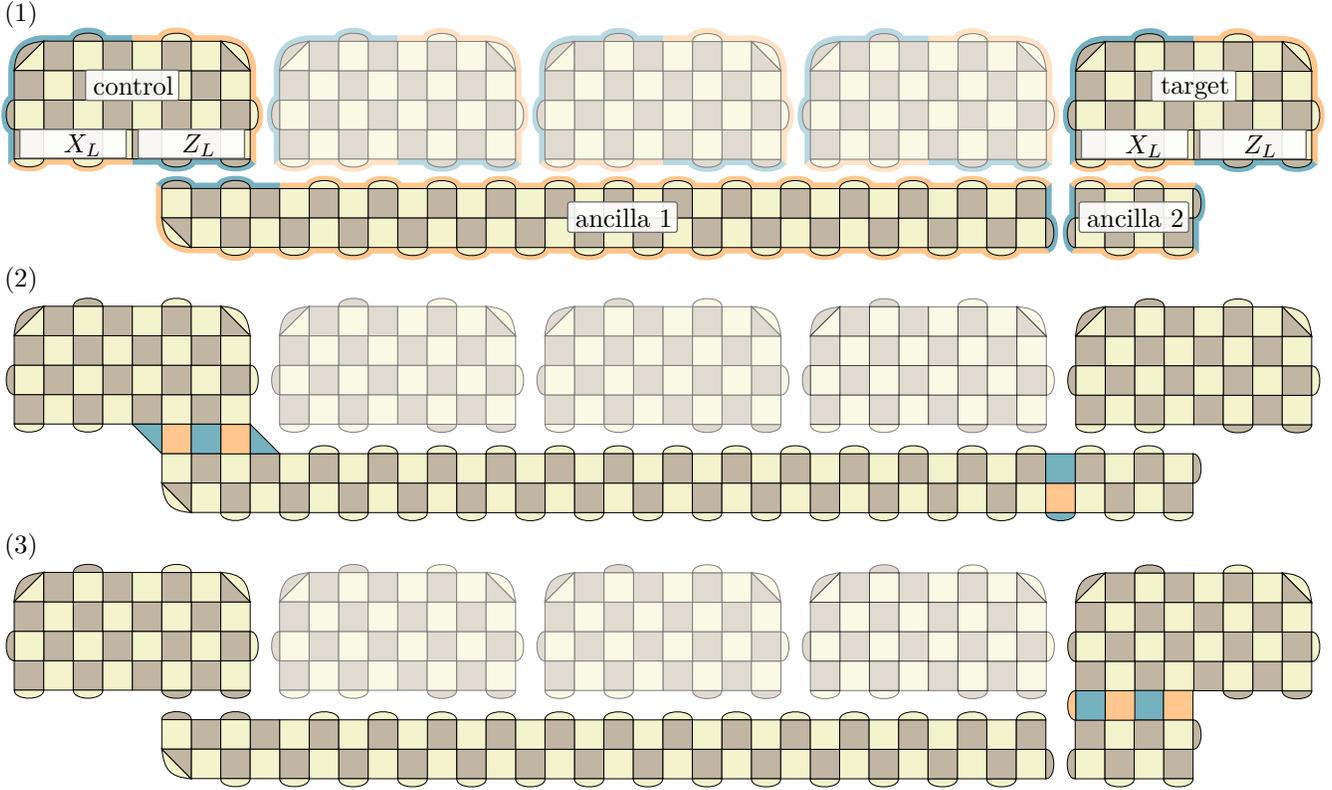
\caption{Long-range CNOT between two wide qubits in the default encoding that are separated by three other qubits. After initializing two ancillas in the $\ket{+}$ state $(1)$, lattice surgery $(2)$ simultaneously measures the $Z_L \otimes Z_L$ parities between control and ancilla 1, and ancilla 1 and ancilla 2. This also yields the $Z_L \otimes Z_L$ parity between control and ancilla 2, such that ancilla 2 can be used for an $X_L \otimes X_L$ parity measurement $(3)$ with the target qubit. At the end of the CNOT protocol, ancilla 1 is read out in the $X$ basis with outcome $m$, leading to a $Z^m$ correction on the control.}
\label{fig:longrange}
\end{figure*}

Let us first discuss standard lattice surgery between two neighboring wide qubits in the default encoding. Consider the CNOT protocol in Fig.~\ref{fig:neighborcnot}. Lattice surgery~\cite{Horsman2012} is a protocol for fault-tolerant logical parity measurements which only requires the measurement of local stabilizer operators. After initializing an ancilla qubit in the $\ket{+}$ state, lattice surgery between the $Z$ edges of the control and ancilla qubit in step $(2)$ measures their $Z_L\otimes Z_L$ parity. This is done by modifying the stabilizers along the boundaries. The boundary $X$ stabilizers are merged to form four-qubit stabilizers (orange), and new $Z$ stabilizers (blue) are introduced. While the stabilizers still mutually commute, this procedure increases the total number of stabilizers by one. In other words, the number of degrees of freedom is reduced by one, and one bit of information is measured during this protocol. The measurement outcome of the orange stabilizers is trivial, as they are products of previously known boundary stabilizers. The outcome of the blue stabilizers, on the other hand, is nontrivial. They contain each boundary qubit exactly once. Therefore, their product is precisely the operator $Z_{\rm edge}^{\rm (control)} \otimes Z_{\rm edge}^{\rm (ancilla)}$, which corresponds to the $Z_L \otimes Z_L$ parity in the default encoding. Thus, lattice surgery implements a fault-tolerant parity measurement between logical qubits. Similarly, in the following lattice surgery step $(3)$, the blue stabilizers are trivial, and the product of orange stabilizers is $X_{\rm edge}^{\rm (ancilla)} \otimes X_{\rm edge}^{\rm (target)}$. A $Z_L$ basis measurement of the ancilla qubit completes the gate circuit in Fig.~\ref{fig:cnotcircuit}. The subsequent Pauli corrections are Clifford gates and can be handled by edge tracking.

\subsection{Long-range CNOT}

\begin{figure*}[t!]
\centering
\def\svgwidth{\linewidth}
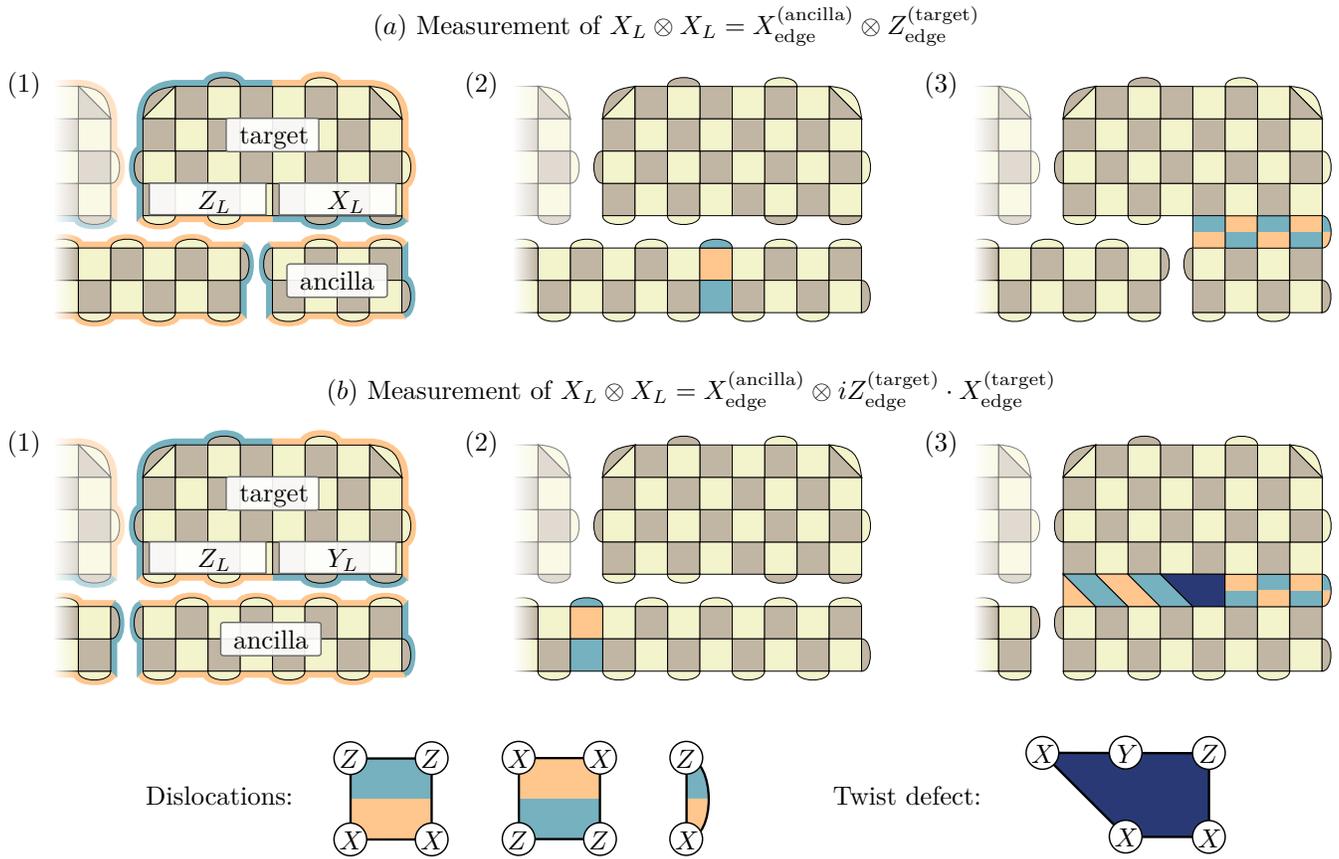
\caption{$X_L \otimes X_L$ parity measurements between an ancilla and an edge-tracked target qubit. In $(a)$, edge tracking has changed the encoding of the target to $X_{\rm edge} = Z_L$ and $Z_{\rm edge} = X_L$. The stabilizer configuration that measures the $X_L \otimes X_L$ parity corresponds to a dislocation line. In $(b)$, the encoding of the target qubit is $X_{\rm edge} = Z_L$ and $Z_{\rm edge} = Y_L$. Here, the $X_L \otimes X_L$ parity is measured by a stabilizer configuration that corresponds to a dislocation line that is terminated by a twist defect.}
\label{fig:twistsurgery}
\end{figure*}

A similar protocol can be used to perform CNOTs between logical qubits that are not nearest neighbors, but separated by some distance. For this, we use lattice surgery to measure the $Z_L \otimes Z_L$ parities between the control qubits and multiple ancilla qubits simultaneously~\cite{Litinski2017,Litinski2017a,Horsman2012}. In the protocol in Fig.~\ref{fig:longrange}, two ancilla qubits are initialized in the $\ket{+}$ state, one long ancilla that spans the entire distance between the control and target, and another that is adjacent to the $X$ edge of the target. In step $(2)$, lattice surgery simultaneously measures the $Z_L \otimes Z_L$ parities between control and long ancilla, and between both ancillas. This effectively measures the $Z_L \otimes Z_L$ parity between control and ancilla 2 as the product of both measurements. Thus, ancilla 2 can be used as the ancilla of the CNOT protocol of Fig.~\ref{fig:cnotcircuit}. An $X_L \otimes X_L$ parity measurement between ancilla 2 and the target qubit, and a subsequent $Z$ basis readout of ancilla 2 complete the CNOT protocol. Since ancilla 1 is still entangled with the control qubit, it cannot be discarded right away, but needs to be measured in the $X$ basis with outcome $m \in \{0,1\}$, which leads to a subsequent $Z^m$ Pauli correction on the control qubit.

Vertical $X$ error strings connecting the (orange) $X$ edges of the long ancilla qubit can introduce errors to the CNOT protocol. While the number of possible error strings increases linearly with the control-target separation $s$, the probability of error strings decreases exponentially with the width of the ancilla. Therefore, the width needs to increase with $\mathcal{O}(\log s)$ in order to maintain the CNOT gate fidelity, implying a space overhead of $\mathcal{O}(s \log s)$ for the long-range CNOT. There are two factors that contribute to the time overhead of the protocol: decoding and syndrome readout errors. While decoding can be done with a runtime that scales with $\mathcal{O}(\log s)$~\cite{Duclos2010}, the correction of stabilizer measurement errors is handled by recording multiple rounds of syndrome extraction for one code cycle~\cite{Dennis2002}. This effectively introduces a third dimension to the code. The number of recorded measurement rounds for each code cycle depends on the measurement fidelity. With higher measurement fidelity, fewer measurement rounds are required to reach the same logical CNOT gate fidelity. As with the width of the long ancilla, errors in the time dimension are suppressed exponentially with the number of measurement rounds, i.e., with the code distance in time, but the number of possible error strings increases linearly with $s$. This implies that the number of measurement rounds needs to increase with $\mathcal{O}(\log s)$. Thus, the total time overhead is still just $\mathcal{O}(\log s)$, which is essentially constant for finite-size systems.

Note that in our figures (such as Fig.~\ref{fig:longrange}), the widths of the ancilla qubits, and therefore their code distances, are chosen to be smaller than the code distances of the wide qubits. This may be a valid choice for some computations, since the ancillas only need to survive for the duration of the CNOT, as opposed to data qubits that may need to survive for the entire computation. In practice, however, we expect that the space reserved for ancilla qubits will be in use for various CNOT gates for essentially the entire duration of the quantum computation. Therefore, for most applications, the code distances of the ancilla qubits and the data qubits should be chosen to be equal, and the logarithmic space overhead scaling with the control-target separation can be ignored. In this case, all logical qubits are protected against error strings of length $(d - 1)/2$ during each code cycle. There is still a logarithmic space overhead scaling, since the necessary code distance to reach a certain target error probability at the end of a quantum computation involving $n$ logical qubits scales with $\mathcal{O}(\log n)$.

\begin{figure*}[t!]
\centering
\def\svgwidth{\linewidth}
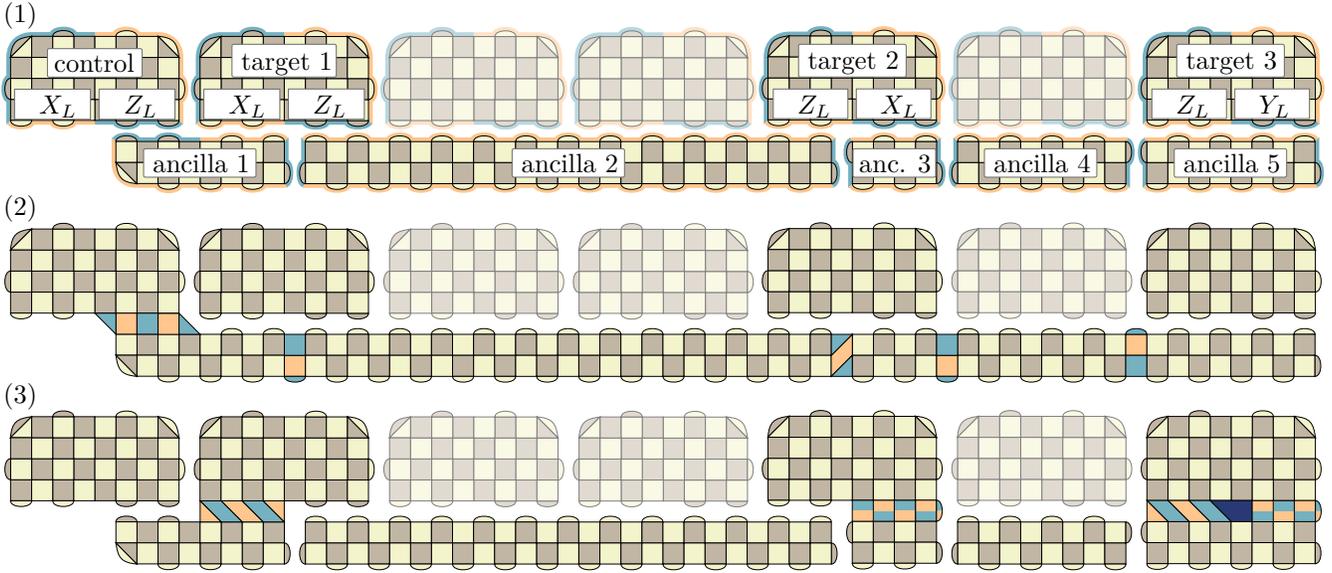
\caption{Long-range multi-target CNOTs with edge-tracked qubits. The control, ancillas, and target 1 are in the default encoding $Z_{\rm edge} = Z_L$ and  $X_{\rm edge} = X_L$, whereas target 2 and target 3 have been modified by edge tracking, such that $X_L \otimes X_L$ parity measurements require lattice surgery between different edge types. Five ancilla qubits are initialized in the $\ket{+}$ state (1) and their $Z_L \otimes Z_L$ parities with the control qubit are measured simultaneously (2). Ancillas 2 and 4 merely provide long-range communication and are not used for CNOTs, but are instead read out in the $X$ basis. Subsequent $X_L \otimes X_L$ parity measurements (3) use the previously discussed lattice surgery protocols for edge-tracked qubits.}
\label{fig:multitarget}
\end{figure*}

\subsection{CNOT between edge-tracked qubits}

The previously discussed standard lattice surgery protocols can be used to measure $Z_{\rm edge} \otimes Z_{\rm edge}$ and \linebreak $X_{\rm edge} \otimes X_{\rm edge}$. However, CNOTs between edge-tracked qubits may require additional parity measurements. This is where dislocations and twist defects come into play.

In Fig.~\ref{fig:twistsurgery}, we explore the two additional situations that may occur for $X_L \otimes X_L$ parity measurements between an ancilla and an edge-tracked target qubit during a CNOT protocol. In the first situation $(a)$, the $X_L$ operator is defined by the target's $Z$ edge as a consequence of edge tracking. Thus, lattice surgery needs to measure the operator $X_{\rm edge}^{\rm (ancilla)} \otimes Z_{\rm edge}^{\rm (target)}$. For this, the boundary stabilizers are merged, and new stabilizers are introduced. One can check that all stabilizers commute, and that the product of the nontrivial stabilizers indeed yields $X_L \otimes X_L$.

The remaining possibility is that, as a consequence of edge tracking, none of the edges of the target define its $X_L$. In $(b)$, the target qubit is in the encoding where $X_{\rm edge} = Z_L$ and $Z_{\rm edge} = Y_L$. Since $X_L = i Z_L Y_L$, and therefore $X_L = i X_{\rm edge} Z_{\rm edge}$, lattice surgery now needs to measure $X_{\rm edge}^{\rm (ancilla)} \otimes \ iX_{\rm edge}^{\rm (target)} \cdot Z_{\rm edge}^{\rm (target)}$. Similar to the previous cases, stabilizers along the boundary in $(b3)$ are merged yielding the trivial stabilizers. The product of the newly introduced nontrivial stabilizers is again the $X_L\otimes X_L$ parity. Note that the center qubit of the blue five-qubit operator contributes to the stabilizer measurement in the $Y$ basis, since it is part of both the $X$ and the $Z$ edge.

The three different lattice surgeries in panel (3) of Fig.~\ref{fig:longrange}, and panels $(a3)$ and $(b3)$ of Fig.~\ref{fig:twistsurgery} can also be interpreted as protocols to measure $X_L \otimes X_L$, $Z_L \otimes X_L$ and $Y_L \otimes X_L$ between a wide qubit and a square qubit in the default encoding. The protocol involving $Y_L$ is what we refer to as twist-based lattice surgery, since the five-qubit operator corresponds to a twist defect.

Such a parity measurement can also be used to measure the product $iX_{\rm edge} \cdot Z_{\rm edge}$ of a qubit, e.g., to read out the qubit in the $Y_L$ basis in the default encoding. For this, an ancilla can be initialized in the $\ket{0}$ state, such that a $Y_L^{\rm (qubit)} \otimes Z_L^{\rm (ancilla)}$ parity measurement between qubit and ancilla is equivalent to a $Y_L$ measurement of the qubit.

This covers all the necessary lattice surgery protocols for CNOTs between edge-tracked qubits. The $Z_L \otimes Z_L$ parity measurements between ancilla qubits and edge-tracked control qubits are analogous to the $X_L \otimes X_L$ parity measurements in Fig.~\ref{fig:twistsurgery}. The concrete implementation of the required stabilizer measurements depends on the given architecture. While Majorana-based implementations allow for direct measurements of the necessary operators, non-topological setups such as superconducting qubits require the use of measurement qubits. In the latter case, the stabilizer measurement protocol requires special care in order to avoid correlated errors that lower the effective code distance, as we show in Appendix~\ref{app:readout}.

\begin{figure*}[t!]
\centering
\def\svgwidth{\linewidth}
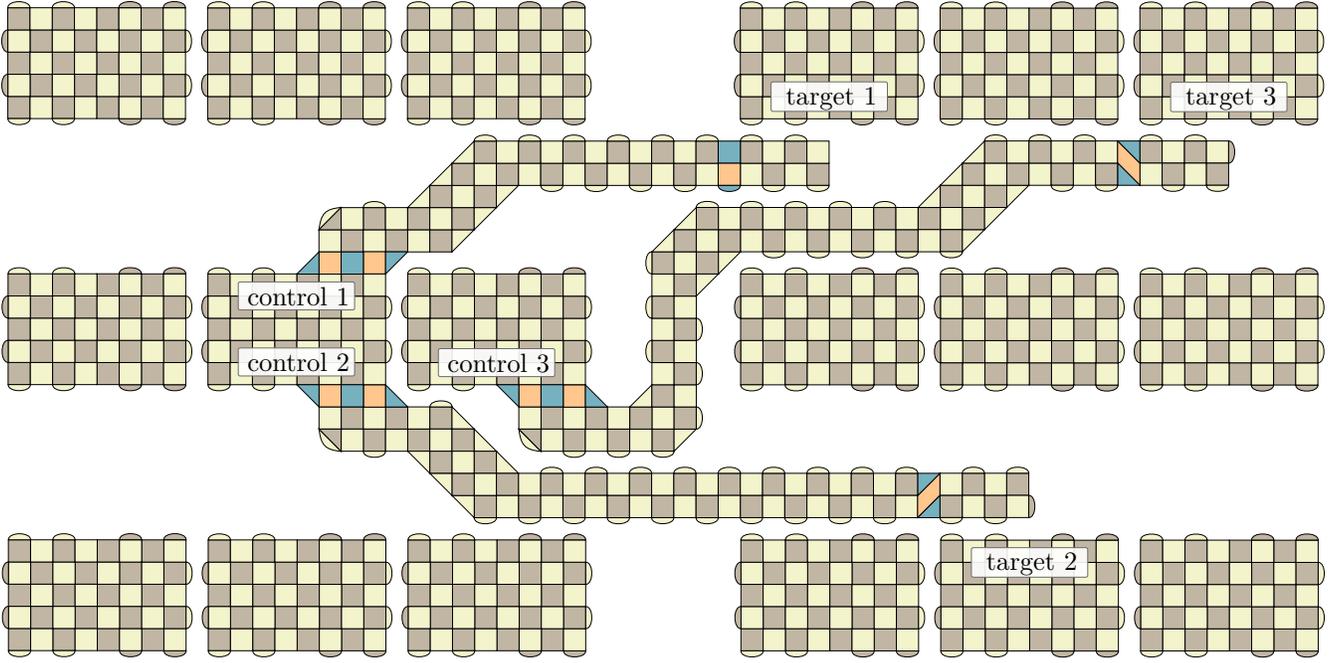
\caption{Example of a two-dimensional arrangement of surface code qubits, where qubits are grouped in blocks of six. The long ancilla qubits can be used for three simultaneous long-range CNOT gates.}
\label{fig:2darr}
\end{figure*}

\subsection{Connection to twist defects}

The stabilizer configurations in these modified lattice surgery protocols feature dislocations and twist defects. The mixed stabilizers in $(a3)$ correspond to a dislocation in the surface code. The stabilizer configuration in $(b3)$ corresponds to a dislocation line between the $X$ edge of the ancilla and the $Z$ edge of the target which is terminated by a five-qubit twist defect~\cite{Bombin2010,Brown2017}.  

Twist-based lattice surgery can also be interpreted in a Majorana fermion picture. It was pointed out that the corners of square surface code qubits (as in Fig.~\ref{fig:qubit}) correspond to twist defects~\cite{Brown2017}. Similarly, the ends of the $X$ and $Z$ edges of wide qubits can be replaced by twist defects~--~i.e., Majorana fermions~--~such that the logical operators $X_L$, $Z_L$, and $Y_L$ are two-Majorana fermion parity operators. Lattice surgery then effectively implements a four-Majorana fermion parity measurement~\cite{Brown2017}. In Fig.~\ref{fig:twistsurgery} $(b3)$, these four Majorana fermions are in the bottom left and right corners of the target, and in the top left and right corners of the ancilla. The twist defect corresponds to the remaining Majorana fermion residing between the $X$ and $Z$ edge of the target qubit, which is not part of the parity measurement.

\subsection{Long-range multi-target CNOT}

The simultaneous $Z_L \otimes Z_L$ parity measurements of long-range CNOTs can be used for multi-target CNOTs, i.e., for multiple CNOTs with the same control, but different target qubits. An example is shown in Fig.~\ref{fig:multitarget}, where five ancillas are used to perform three CNOTs with three edge-tracked targets simultaneously. Step $(2)$ shows the simultaneous measurement of $Z_L \otimes Z_L$ parities of six neighboring qubits, which correspond to one control and five ancilla qubit. This protocol effectively measures the $Z_L \otimes Z_L$ parities of all pairs of qubits, and in particular of the control and each ancilla qubit. Thus, each of the five ancilla qubits can be used for $X_L \otimes X_L$ parity measurements with target qubits. While ancillas 1, 3, and 5 are used for CNOTs with targets 1, 2, and 3, ancillas 2 and 4 merely bridge distances between distant qubits. 

\begin{figure}[b!]
\centering
\def\svgwidth{\linewidth}
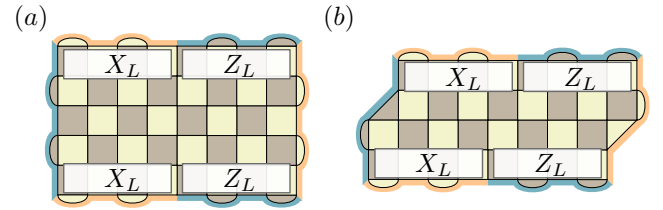
\caption{Double-sided qubits encode two logical qubits using $(a)$ $2d^2+d-1$ or $(b)$ $2d^2-d$ physical qubits. The left and right edges correspond to the logical operators $Z_L \otimes Z_L$ and $X_L \otimes X_L$ of both encoded qubits, respectively.}
\label{fig:doublesided}
\end{figure}

\begin{figure*}[t!]
\centering
\def\svgwidth{\linewidth}
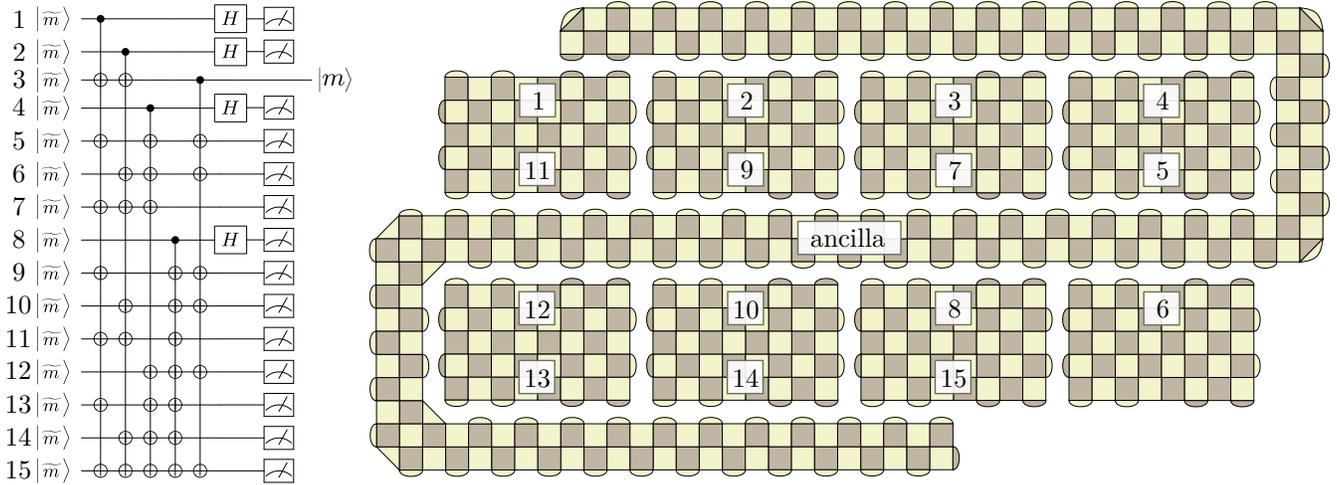
\caption{Example of the 15-to-1 magic state distillation protocol using long-range multi-target CNOTs via lattice surgery. By appropriately partitioning the long ancilla qubit, it can be used for each of the five multi-target CNOTs of the protocol.}
\label{fig:distillation}
\end{figure*}

Thus, by simultaneously initializing multiple ancillas, lattice surgery provides long-range multi-target CNOTs with edge-tracked qubits with the same time overhead as single CNOTs. At the end of the protocol, ancillas that are used for $X_L \otimes X_L$ parity measurements with target qubits are read out in the $Z$ basis, whereas ancillas used to bridge long distances are read out in the $X$ basis. Multi-target CNOTs are particularly useful for logical $T$ gates, as magic state distillation schemes typically consist of many multi-target CNOTs. These complete the universal gate set of our scheme, as we discuss in the following section.

\section{2D arrangement of logical qubits}
\label{sec:2d}

So far, we have considered logical qubits arranged on a line. The lattice-surgery-based CNOT gates can also provide long-range connectivity in two dimensions. For this, it will be convenient to use the space of wide qubits to encode \textit{two} logical qubits instead of just one. The double-sided qubits shown in Fig.~\ref{fig:doublesided} reduce the space overhead from $\sim 2d^2$ physical qubits for each logical (wide) qubit back to $\sim d^2$ physical qubits, similar to the square qubits in Fig.~\ref{fig:qubit}. The downside of double-sided qubits is that state initialization and readout is more complicated, as the two encoded qubits cannot be measured independently. However, one can use lattice surgery to initialize and read out in any Pauli basis. For instance, a qubit can be initialized in the $\ket{0}$ state by initializing a standard ancilla encoding a single qubit in the $\ket{0}$ state and performing lattice surgery via the $Z$ edges of both qubits. Readout is done the same way, using the appropriate edge of the qubit. Should one require fast initialization and readout, it is still possible to use wide qubits instead of double-sided qubits.

An example of a 2D arrangement of double-sided qubits is shown in Fig.~\ref{fig:2darr}, where they form blocks of six logical qubits. The space between blocks is used for ancilla qubits for long-range CNOT gates. The separation between blocks not only sets the maximum width of the ancilla qubits, but also influences the number of multi-target CNOTs that can be performed simultaneously. The larger the separation, the more ancilla qubits can fit between the qubit blocks. The example in Fig.~\ref{fig:2darr} shows three simultaneous CNOT gates, where the space between qubit blocks allows for two parallel ``lanes'' of CNOT ancillas.
Thus, a larger separation between qubit blocks increases the connectivity, but also the space overhead.

\subsection{Example: Magic state distillation}
Having discussed the implementation of the logical Clifford gates in our scheme, the remaining gate for universal quantum computation is the logical $T$ gate. One possibility to implement the logical $T$ gate using physical $T$ gates and logical Clifford gates is magic state distillation~\cite{Bravyi2005}. The aim of this scheme is to generate an encoded magic state $\ket{m} = (\ket{0} + e^{i \pi/4}\ket{1})/\sqrt{2}$, which corresponds to a $\ket{+}$-state on which a $T$ gate has been performed. A CNOT gate between $\ket{m}$ and a target qubit, followed by the measurement of $\ket{m}$ corresponds to a logical $T$ gate on the target qubit, up to a Clifford correction.

\begin{table*}[t!]
\centering
\begin{tabular}{c|c|c|c}
& color code & wide surface code & double-sided surface code \\ \hline 
space overhead & +~low ($\approx \frac{3}{4}d^2$ or $\frac{1}{2}d^2$) & --~high ($\approx 2d^2$)& $\sim$~moderate ($\approx d^2$) \\ \hline
initialization \& readout & +~fast $X,Y,Z$ & $\sim$~fast $X,Z$; slow $Y$ & --~slow $X,Y,Z$ \\ \hline
stabilizer weight & --~high (six or eight) & +~low (four) & +~low (four)
\end{tabular}
\caption{Comparison between color-code-based~\cite{Landahl2011,Litinski2017a} and surface-code-based schemes.}
\label{tab:comp}
\end{table*}

However, it is only possible to prepare \textit{physical} magic states, which are moreover faulty states $\ket{\widetilde{m}}$, i.e., generated using an imprecise physical $T$ gate. These physical states can be converted into \textit{logical} faulty magic states $\ket{\widetilde{m}}$ via state injection~\cite{Horsman2012}. Magic state distillation protocols take many faulty magic states and convert them to fewer, but more precise magic states. These protocols typically consist of many multi-target CNOT gates.

One example of a magic state distillation protocol is shown in Fig.~\ref{fig:distillation} for the example of 15-to-1 conversion~\cite{Bravyi2005}, which converts 15 faulty magic states into one better magic state. It consists of 34 CNOT gates grouped into five multi-target CNOTs. The figure also shows an arrangement of qubits that can be used to implement the protocol. By appropriately partitioning the long ancilla qubit, each of the five multi-target CNOTs can be performed using the protocol in Fig.~\ref{fig:multitarget}. We provide the detailed stabilizer configurations for this \linebreak 15-to-1 conversion in Appendix \ref{app:distillation}.
The space overhead of the 15-to-1 conversion depends on the code distances of the magic states and the width of the ancilla. The time overhead is mostly determined by the five multi-target CNOTs, which require two code cycles (including repetitions accounting for stabilizer measurement errors) for their parity measurements by lattice surgery.

\section{Conclusion}

\renewcommand{\arraystretch}{1.5}

We have demonstrated that edge tracking can be used to eliminate the time overhead of logical single-qubit Clifford gates in surface codes, as should be expected considering the Gottesman-Knill theorem. Twist-based lattice surgery provides long-range multi-target CNOTs with a time overhead that only scales with $\mathcal{O}(\log s)$ of the control-target separation $s$, and a space overhead that scales with $\mathcal{O}(s \log s)$. Compared to color code qubits, the surface code qubits used in our scheme require more physical qubits ($\sim d^2$) for each logical qubit with code distance $d$, but~--~with the exception of twist defects~--~only require the measurement of weight-four stabilizers. Our scheme can provide full 2D connectivity between logical qubits, where the degree of connectivity is governed by the separation of qubit blocks, and therefore by the space overhead. Together with magic state distillation, our scheme allows for fault-tolerant universal quantum computation.

\begin{figure}[b!]
\centering
\def\svgwidth{\linewidth}
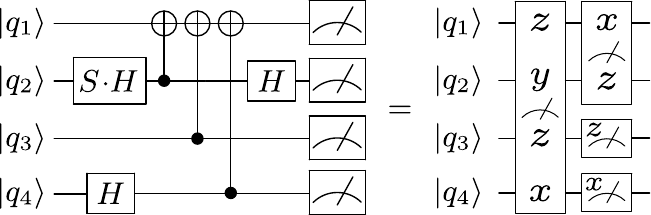
\caption{Example of a Clifford circuit that is reduced to Pauli product measurements.}
\label{fig:trackingcircuit}
\end{figure}

One may be wondering whether there is still any advantage offered by the transversal single-qubit Clifford gates of color codes and the color-code-based lattice-surgery scheme presented in Ref.~\cite{Litinski2017a}. A comparison of these codes is shown in Tab.~\ref{tab:comp}. While color codes require the measurement of higher-weight stabilizers, they offer fast qubit readout in all Pauli bases, and a lower space overhead of $\sim \frac{3}{4}d^2$ physical qubits per logical qubit for 6.6.6 color codes, or even $\sim \frac{1}{2}d^2$ for 4.8.8 color codes. So if the measurement of higher-weight stabilizers is not substantially more difficult in a given physical implementation, as might be the case for Majorana-based qubits, it is advantageous to use the color-code-based scheme. In other implementations, such as superconducting qubits, the difficulty of higher-weight stabilizer measurements shifts the preference towards surface-code-based architectures.

\begin{figure}[b!]
\centering
\def\svgwidth{\linewidth}
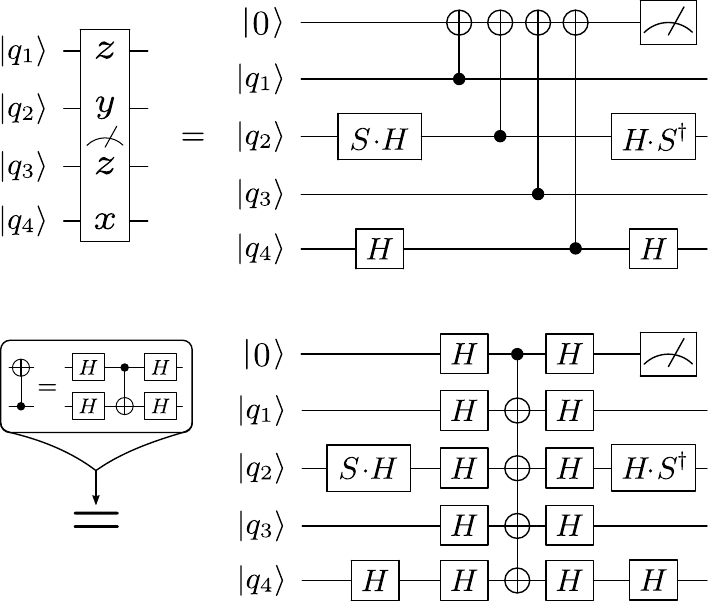
\caption{Circuit identity for the measurement of the Pauli product operator $Z \otimes Y \otimes Z \otimes X$ using an ancilla and a multi-target CNOT gate. The circuit identity exploits the fact that the roles of control and target can be reversed by the application of Hadamard gates before and after a CNOT gate. Any product of Pauli operators can be measured this way.}
\label{fig:pauliprod}
\end{figure}

\begin{figure*}[t!]
\centering
\def\svgwidth{\linewidth}
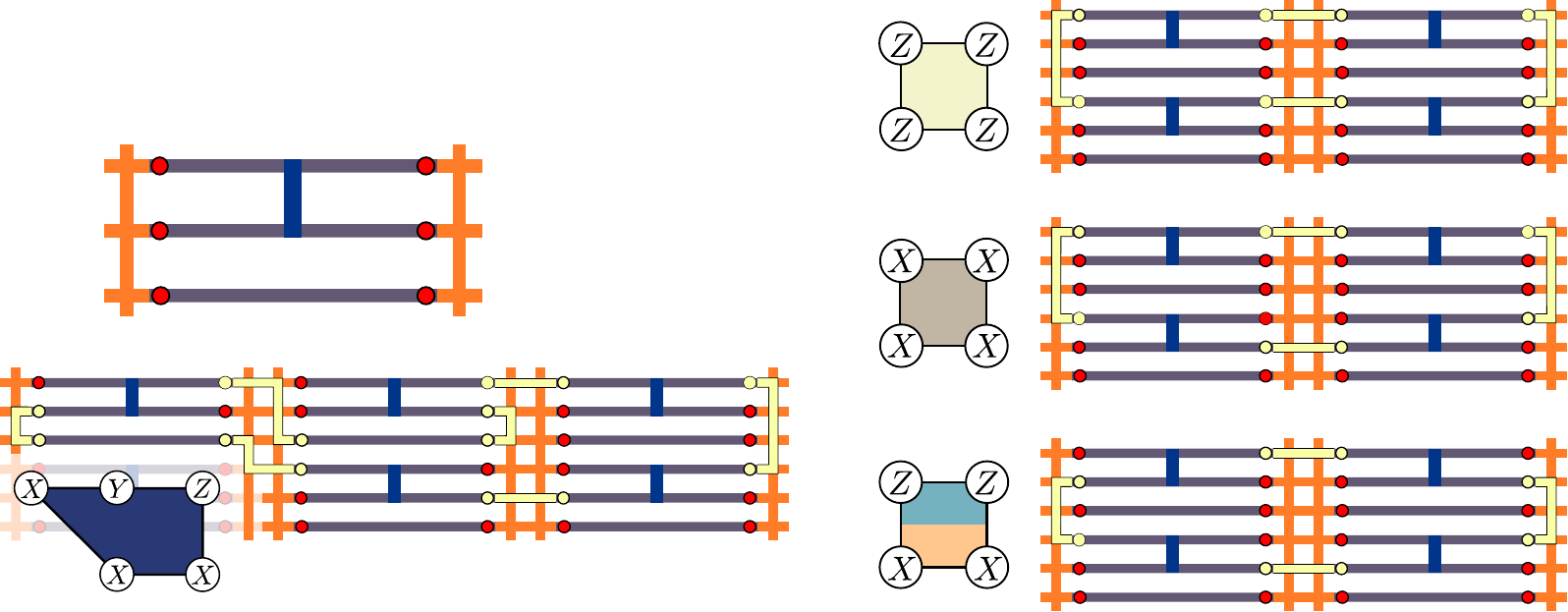
\caption{Tunnel coupling configurations for the measurement of various check operators using a square network of Majorana-based tetron qubits, as introduced in Ref.~\cite{Karzig2016}.}
\label{fig:tetron}
\end{figure*}

An important point is that the Gottesman-Knill theorem allows for the classical tracking of all Clifford gates, including CNOT gates.
As CNOT gates map $X \otimes \mathbbm{1} \rightarrow X \otimes X$ and $\mathbbm{1} \otimes Z \rightarrow Z \otimes Z$, tracking of CNOTs does not preserve the locality of the logical operators, in contrast to single-qubit Clifford gates. By tracking \textit{all} Clifford gates, any layer of Clifford gates followed by $n$ single-qubit measurements can always be compressed to $n$ measurements of \textit{nonlocal} products of Pauli operators without any preceding gate operations (see Fig.~\ref{fig:trackingcircuit} for an example). With distilled magic states as a resource, any non-Clifford gate corresponds to trackable Clifford gates and a measurement of the magic state. In this case, Pauli product measurements are the only hardware operations that need to be performed explicitly. The fault-tolerant measurement of any nonlocal Pauli product can be implemented using an ancilla qubit and a multi-target CNOT gate on edge-tracked qubits. An example of such a protocol is shown in Fig.~\ref{fig:pauliprod} for the measurement of the Pauli product $Z_L \otimes Y_L \otimes Z_L \otimes X_L$. Thus, any quantum computation can be performed using only two types of hardware operations: distillation of resource states and Pauli product measurements via multi-target CNOT gates on edge-tracked qubits.

A crucial problem of quantum information theory is the optimization of quantum circuits in order to minimize the space-time overhead of any quantum computation. However, any circuit optimization depends on the constraints set by the quantum computer hardware and the code used for error correction. Based on the existing schemes for surface-code and color-code quantum computation, the following minimal assumptions concerning the underlying hardware and the logical operations accessible by the code appear reasonable: $(i)$ The underlying hardware can measure local products of physical Pauli operators.  $(ii)$ The quantum error-correcting code allows for the measurement of nonlocal products of logical Pauli operators. $(iii)$ Resource states can be generated for the implementation of logical non-Clifford gates. Based on these constraints, an important circuit optimization problem is to find heuristics that minimize the number of required resource states and the number of layers of Pauli product measurements, as these are the only operations that cannot be relegated to a classical computer.

Open questions related to our surface-code scheme include the efficient decoding of wide, long and double-sided qubits, estimations of their survival times, and implementations of our scheme in a concrete physical system. Our scheme may also be adapted to surface-code quantum computing with twist-based triangle codes~\cite{Yoder2017}, in order to avoid the reorientation of triangles, and to further reduce the space overhead of surface codes. We hope that our lattice-surgery-based approach can contribute to ongoing efforts to realize a surface-code quantum computer.

\section*{Acknowledgments}
We thank Benjamin J.~Brown, Jens Eisert, Markus S.~Kesselring, and Fernando Pastawski for insightful discussions. This work has been supported by the Deutsche Forschungsgemeinschaft (Bonn) within the network CRC TR 183.

\appendix

\section{Stabilizer measurements in concrete implementations}
\label{app:readout}

Our twist-based surgery scheme requires the measurement of certain operators that are products of Pauli operators on up to 5 qubits. How these operators are measured in practice depends on the concrete hardware used for quantum computing. In this appendix, we show how these measurements could in principle be implemented with Majorana-based qubits, and with non-topological qubits such as superconducting qubits that require the use of ancillary measurement qubits for stabilizer readout.

\begin{figure}
\centering
\def\svgwidth{\linewidth}
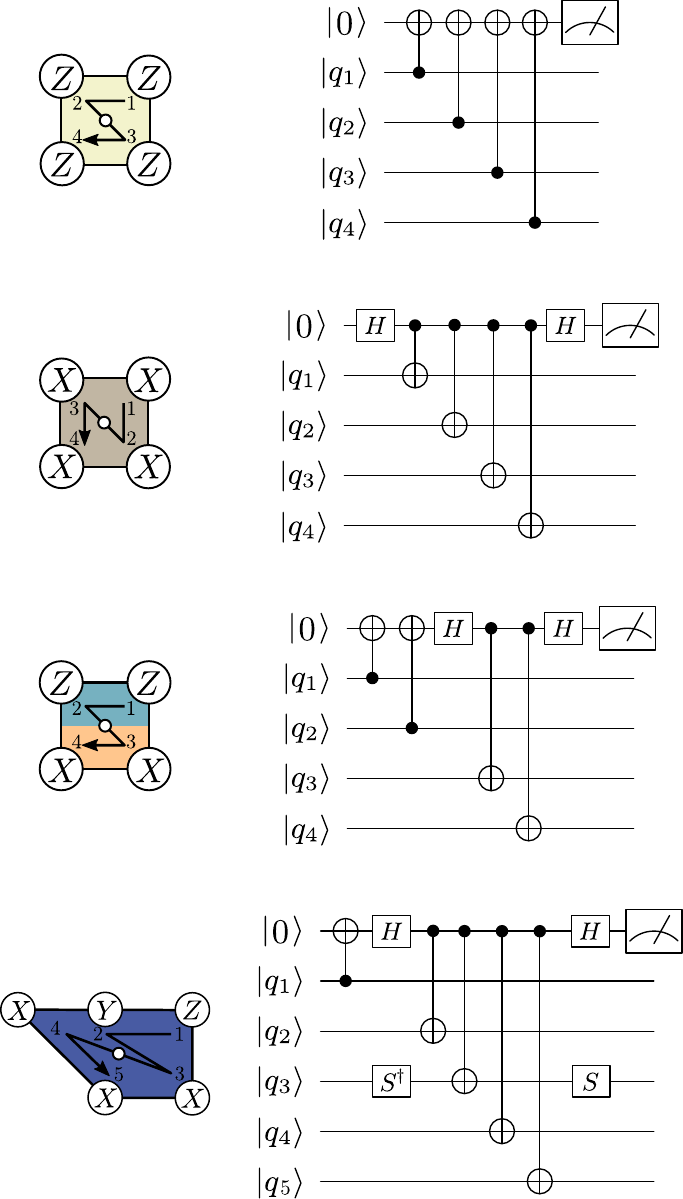
\caption{Circuits for the stabilizer readout using an ancillary measurement qubit placed in the center of the stabilizer.}
\label{fig:readoutcircuits}
\end{figure} 

\subsection{Majorana-based qubits}

The primary operation of Majorana-based qubits is the measurement of local products of Majorana operators, which correspond to local products of Pauli operators. Thus, they can be straightforwardly used to measure the stabilizers in our twist-based surgery scheme. In Fig.~\ref{fig:tetron}, we show how this can be done in a network of tetron qubits introduced in Ref.~\cite{Karzig2016}. In a nutshell, these are qubits that are encoded in the doubly degenerate ground-state space of four Majorana zero modes $\gamma_1 \dots \gamma_4$ that are localized at the ends of two topological superconducting nanowires which are put into a fixed parity sector ($\gamma_1\gamma_2\gamma_3\gamma_4 = -1$) by a non-topological superconductor bridging the two wires. The Majorana operators are self-conjugate $\gamma = \gamma^\dagger$ and mutually anticommute $\{\gamma_i,\gamma_j\} = 2\delta_{i,j}$. Therefore, the Pauli operators of each tetron qubit can be chosen as $Z=i\gamma_1\gamma_2$ and $X=i\gamma_2\gamma_3$.

In a square lattice of tetrons, each tetron qubit is connected to a network of semiconductors. Local products of Majorana operators are measured by opening tunnel couplings between tetrons and the semiconductor network, such that the tunnel couplings form closed paths. The semiconducting wire segments between tetrons form quantum dots whose energy levels are shifted by virtual processes that tunnel electrons around the closed path. Since these processes involve each Majorana operator along the path exactly once, spectroscopy on any of the dots can be used to measure the product of the Majorana operators along the path. In Fig.~\ref{fig:tetron}, we show tunnel coupling configurations that can be used to measure $X$ and $Z$ stabilizers, dislocation operators, and twist operators. For the twist operator, in particular, additional Majoranas $\gamma_a$ and $\gamma_b$ in a fixed parity sector $i\gamma_a\gamma_b=1$ are used as so-called coherent links in order to form the closed path. More details on operator measurements in tetron networks are found in Refs.~\cite{Karzig2016,Litinski2017a}.

\subsection{Non-topological qubits}

For non-topological qubits such as superconducting qubits, Pauli products cannot be measured directly, but are usually read out using ancilla qubits (measurements qubits) that are located in the center of each stabilizer operator, such as in the scheme of Ref.~\cite{Fowler2012}. These measurement qubits are entangled via two-qubit gates with each data qubit that is part of the stabilizer. Afterwards, they are read out to yield the corresponding Pauli product. The readout can be done using the circuits shown in Fig.~\ref{fig:readoutcircuits}. Depending on the elementary operations accessible in a given hardware, a different (but equivalent) circuit may be used, but in any case the readout of each stabilizer requires up to 5 two-qubit gates which need to be performed in succession.

One practical problem of this approach to stabilizer measurements is that due to the use of two-qubit gates, single errors on measurement qubits can spread to multiple data qubits. This can lead to correlated errors which are referred to as hook errors~\cite{Dennis2002}. In particular, $Z$ errors on measurement qubits of $Z$ stabilizers can lead to correlated $Z$ errors on the surrounding data qubits. Similarly, $X$ errors on $X$ measurement qubits lead to correlated $X$ errors. Since three errors are equivalent to just one error (up to a multiplication with a stabilizer), the worst case is the case of one error on a measurement qubit leading to two errors on data qubits. Since these errors will occur on the first two (or last two) qubits that were part of entangling two-qubit gates, the order of the two-qubit gates in the circuits of Fig.~\ref{fig:readoutcircuits} is important.

The aim  is to avoid these correlated errors from lowering the effective code distance. That is, we need to find an ordering of the two-qubit gates, such that a logical operator of a distance $d$ qubit can only be formed by no fewer than $d$ errors. For square surface code patches (as in Fig.~\ref{fig:qubit}), this is done by orienting the ordering of CNOT gates for $X$ stabilizers in an N shape (or in a \reflectbox{N} shape), and for $Z$ stabilizers in a Z shape (or \reflectbox{Z} shape), as was shown in Ref.~\cite{Tomita2014}. This guarantees that correlated $Z$ errors only form in the horizontal direction, while logical $Z$ strings are all oriented vertically. Similarly, correlated $X$ errors form horizontally, which does not contribute towards a vertical logical $X$ string.

\begin{figure}[t]
\centering
\def\svgwidth{\linewidth}
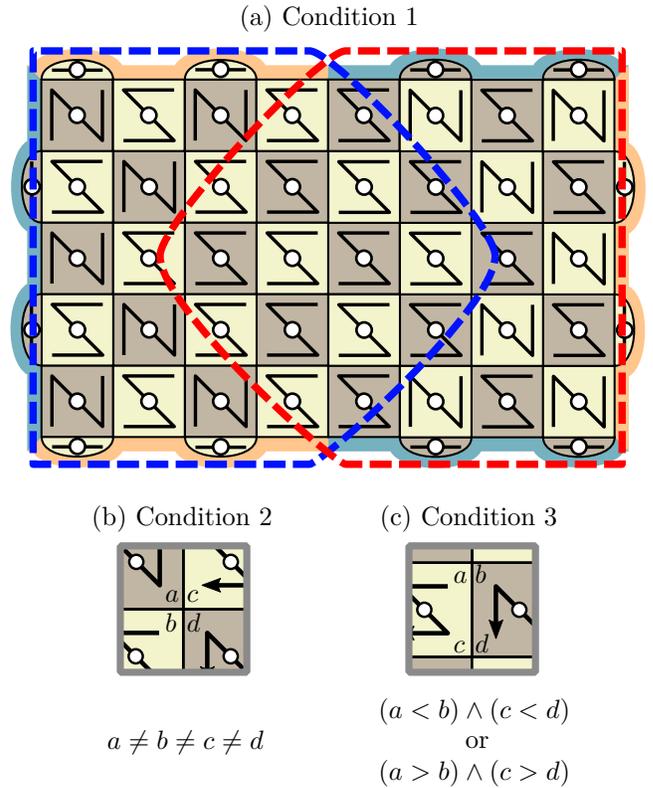
\caption{Three conditions for a valid ordering of two-qubit gates during stabilizer readout. (a) In the blue region, the $Z$ stabilizers are oriented in a Z shape, and outside of this region in an N shape. In the red region, the $X$ stabilizers are oriented in a Z shape, and outside of this region in an N shape. This ensures that hook errors do not lower the effective code distance. (b) The time steps $a$, $b$, $c$, and $d$, that are assigned to the (up to) four two-qubit gates of one data qubits need to be all different. (c) For all edges between neighboring stabilizers, the condition shown in the figure needs to be fulfilled to ensure that the sequence of two-qubit measurements reproduces the desired stabilizer measurements.}
\label{fig:conditions}
\end{figure} 

However, in our scheme, we use the double-sided qubits of Fig.~\ref{fig:doublesided}, which have $Z$ and $X$ operators in both the horizontal and vertical direction. Thus, it is not sufficient to assign one orientation to the $Z$ stabilizers and the other to $X$ stabilizers. In fact, the left half of the double-sided qubit looks like the square qubit in Fig.~\ref{fig:qubit}, i.e., $Z$ stabilizers should be oriented in a Z shape, and $X$ stabilizers in an N shape. In contrast, the right half of the double-sided qubit looks like a rotated square qubit, i.e., $Z$ stabilizer should be oriented in an N shape, and $X$ stabilizers in a Z shape. In the crossover region in the center, both logical $X$ and $Z$ operators are vertical strings, such that both $Z$ and $X$ stabilizers should be oriented in a Z shape. This motivates the first condition for a valid ordering shown in Fig.~\ref{fig:conditions}a. $Z$ stabilizers in the blue region should be oriented in a Z shape, and in an N shape outside of the blue region. $X$ stabilizers in the red region should be oriented in a Z shape, and in an N shape outside of the red region. One can verify that with this choice of orientations, no logical operator string can be formed with fewer than $d$ physical errors.

The only remaining problem is the scheduling of the two-qubit gates. Since the largest check operator is the 5-qubit twist defect, each two-qubit gate needs to be assigned to one of 5 time steps. This implies two other conditions on the ordering of the two-qubit gates. Since each data qubit can only be part of one two-qubit gate in a given time step, the four time steps assigned to the two-qubit gates that a given data qubit is part of need to be all different, which is the second condition in Fig.~\ref{fig:conditions}b.

\begin{figure*}
\centering
\def\svgwidth{0.95\linewidth}
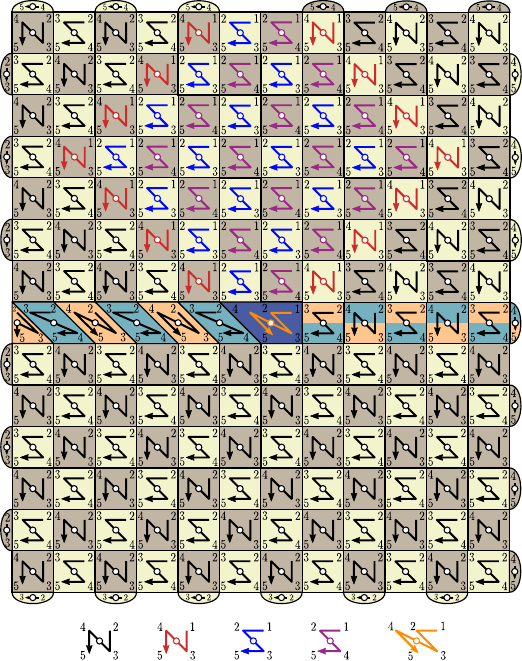
\caption{Possible ordering of two-qubit gates that fulfills the three conditions shown in Fig.~\ref{fig:conditions}. This stabilizer configuration is the most general, as it involves bulk stabilizers, a dislocation line, and a twist defect. It corresponds to a twist-based lattice surgery between a double-sided qubit and a standard rectangular qubit.}
\label{fig:nohooks}
\end{figure*} 

Finally, one needs to ensure that the sequence of two-qubit measurements reproduces the desired stabilizer measurements. For this, consider the action of the CNOT gates of the readout circuit for a $Z$ stabilizer in the Heisenberg picture (or see Appendix B of Ref.~\cite{Fowler2012}). The aim of the readout circuit is to map the $Z$ operator of the measurement qubit onto the operator $Z^{\otimes 5}$ on the measurement qubit and all four data qubits. Since a CNOT maps $\mathbbm{1} \otimes Z$ onto $Z \otimes Z$ (where the first qubit is the control and the second is the target), the circuit in Fig.~\ref{fig:readoutcircuits} achieves exactly that. However, notice that two CNOT gates of the readout circuit of a neighboring $X$ stabilizer also act on the $Z$ operators of the data qubits. These may map the operator onto an operator that involves the $Z$ operator of the wrong measurement qubit, i.e., the measurement qubit of the neighboring stabilizer. This needs to be avoided, as this neighboring measurement qubit is measured in $X$, which anticommutes with the aforementioned operator, leading to random measurement outcomes. For concreteness, let us refer to the time steps assigned to the two CNOTs of the $Z$ stabilizer in question as $a$ and $c$, and to the CNOTs of the $X$ stabilizer as $b$ and $d$, as in Fig.~\ref{fig:conditions}c. There are only two choices of time steps that map the $Z$ operators of the measurement qubit to the correct $Z^{\otimes 5}$ operator. The first possibility is that $a>b$ and $c>d$, such that the two CNOTs of the neighboring $X$ stabilizer have already been performed, which precludes them from acting on the $Z$ operator of the data qubit. The second possibility is that $a<b$ and $c<d$, such that the mapping is performed twice: The first CNOT maps $\mathbbm{1} \otimes Z$ to $Z \otimes Z$, and the second CNOT maps $Z \otimes Z$ back to $\mathbbm{1} \otimes Z$. This is the third condition shown in Fig.~\ref{fig:conditions}c. It needs to hold for all edges between neighboring stabilizers.

It is possible to find an ordering of two-qubit gates in 5 time steps (due to the 5-qubit twist operators) that fulfills all three conditions. Such a possibility is shown in Fig.~\ref{fig:nohooks}. The figure shows the most generic situation which involves the bulk stabilizers of standard and double-sided qubits, as well as a dislocation line and a twist defect.

\section{Magic state distillation protocol}
\label{app:distillation}

Here, we explicitly show the lattice surgery protocols for the multi-target CNOTs part of the 15-to-1 magic state distillation scheme in Fig.~\ref{fig:distillation}. Figures \ref{fig:msd1} and \ref{fig:msd2} show the five multi-target CNOTs of the distillation protocol, where the control and target qubits are highlighted in blue and orange, respectively. Note that the default encodings of the $X$ and $Z$ edges of qubits 5, 9 and 11 are inverted in this protocol. The figures only show the  $Z_L \otimes Z_L$ parity measurements. The subsequent \linebreak  $X_L \otimes X_L$ parity measurements are done via lattices surgeries between the highlighted orange edges and the adjacent ancilla qubits.

\bibliographystyle{apsrev4-1mod}
\bibliography{biblio}

\begin{widetext}

\begin{figure}[H]
\centering
\def\svgwidth{0.68\linewidth}
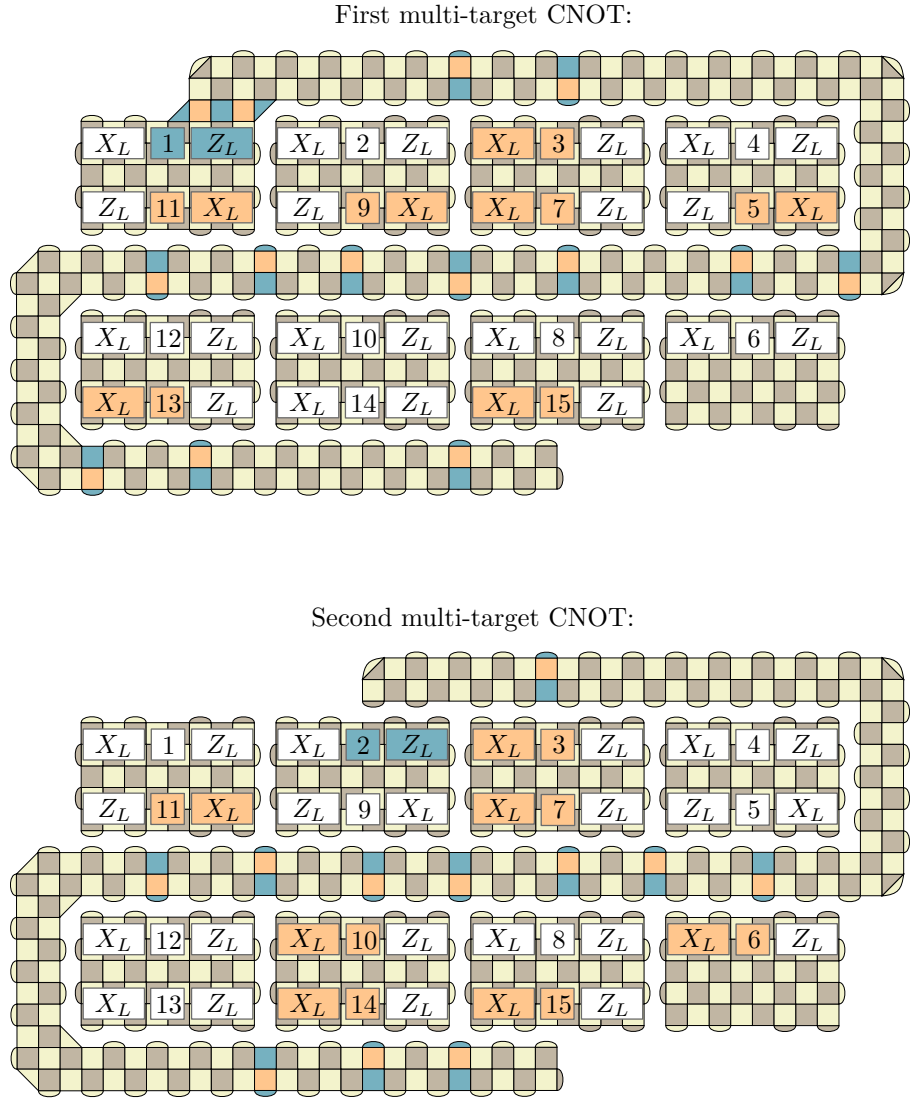
\caption{First and second multi-target CNOT of the distillation protocol in Fig.~\ref{fig:distillation}.}
\label{fig:msd1}
\end{figure} 

\begin{figure}[H]
\centering
\def\svgwidth{0.75\linewidth}
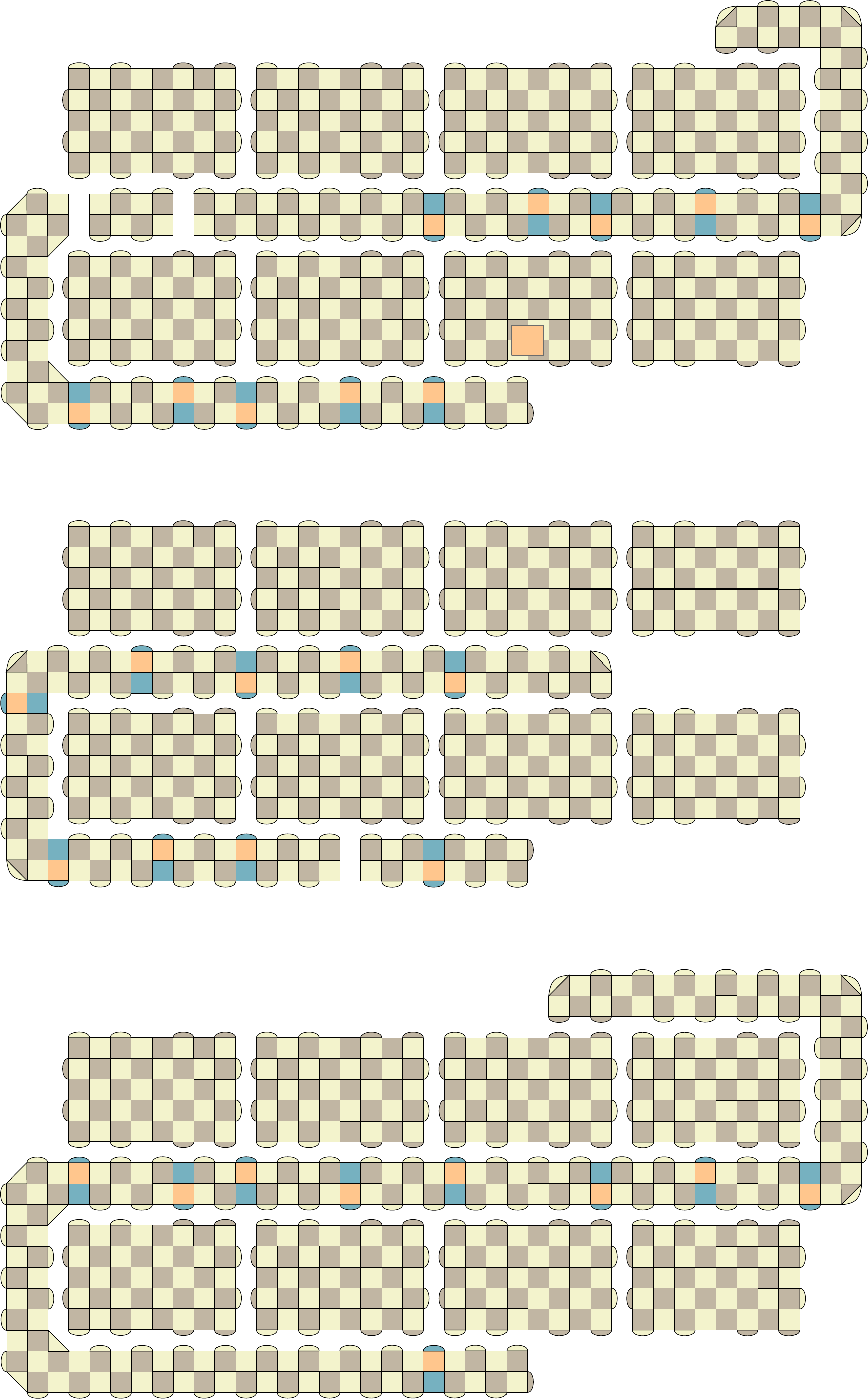
\caption{Third, fourth and fifth multi-target CNOT of the distillation protocol in Fig.~\ref{fig:distillation}.}
\label{fig:msd2}
\end{figure}

\end{widetext}

\end{document}